\documentclass[iop]{emulateapj}

\usepackage{mathptmx}
\usepackage{verbatim}
\usepackage{graphicx}
\usepackage{epstopdf}
\usepackage{amsmath}
\usepackage{textcomp}
\usepackage{latexsym}
\usepackage{multirow}
\usepackage{wasysym}
\usepackage{natbib}
\usepackage{subfigure}
\usepackage{booktabs}

\newcommand{\ltsim}{\raisebox{-.5ex}{$\;\stackrel{<}{\sim}\;$}}
\newcommand{\gtsim}{\raisebox{-.5ex}{$\;\stackrel{>}{\sim}\;$}}
\newcommand{\kms}{\ifmmode {\rm km\ s}^{-1} \else km s$^{-1}$\fi}

\newcommand{\lledd}{$L/L_{\rm Edd}$}

\newcommand{\et}{et al.\ }
\newcommand{\xray}{\hbox{X-ray}}
\newcommand{\aox}{$\alpha_{\rm ox}$}

\newcommand{\nh}{$N_{\rm H}$}

\newcommand{\Ka}{Fe K$\alpha$}
\newcommand{\xmm}{{\sl XMM-Newton}}
\newcommand{\chandra}{{\sl Chandra}}

\newcommand{\lya}{Ly$\alpha$}
\newcommand{\hb}{H$\beta$}
\newcommand{\civ}{C~{\sc iv}}

\journalinfo{The Astrophysical Journal, ???:??? (??pp), 2018 ??? ??}
\slugcomment{Received 2018 April 24; accepted 2018 August 3}

\shortauthors{MARLAR ET AL.}
\shorttitle{X-RAY SPECTRA OF WLQS}

\begin{document}
\let\uppercase\relax \title{Steep Hard-X-ray Spectra Indicate Extremely High Accretion Rates in Weak Emission-Line Quasars}
{\let\thefootnote\relax \footnote{Based on observations obtained with \xmm, an ESA science mission with instruments and contributions directly funded by ESA Member States and NASA.}}
\author{
Andrea~Marlar\altaffilmark{1},
Ohad~Shemmer\altaffilmark{1},
S.F.~Anderson\altaffilmark{2},
W.N.~Brandt\altaffilmark{3,4,5},
A.M.~Diamond-Stanic\altaffilmark{6},
X.~Fan\altaffilmark{7},
B.~Luo\altaffilmark{8,9,10},
R.M.~Plotkin\altaffilmark{11},
Gordon~T.~Richards\altaffilmark{12},
D.P.~Schneider\altaffilmark{3,4},
Jianfeng~Wu\altaffilmark{13}
}
\altaffiltext{1}{
	Department of Physics, University of North Texas, Denton, TX 76203, USA; Andrea.Marlar@unt.edu} 

\altaffiltext{2}{
	Department of Astronomy, University of Washington, Box 351580, Seattle, WA 98195, USA}
	
\altaffiltext{3}{
	Department of Astronomy and Astrophysics, 525 Davey Lab, The Pennsylvania State University, University Park, PA 16802, USA}
	
\altaffiltext{4}{
	Institute for Gravitation and the Cosmos, The Pennsylvania State University, University Park, PA, 16802, USA}
	
\altaffiltext{5}{
	Department of Physics, 104 Davey Lab, The Pennsylvania State University, University Park, PA, 16802, USA}
	
\altaffiltext{6}{
	Department of Physics and Astronomy, Bates College, 44 Campus Avenue, Carnegie Science Hall, Lewiston, ME 04240, USA}
	
\altaffiltext{7}{
	Steward Observatory, University of Arizona, 933 North Cherry Avenue, Tucson, AZ 85721, USA}

\altaffiltext{8}{
	School of Astronomy and Space Science, Nanjing University, Nanjing, Jiangsu 210093, China}
	
\altaffiltext{9}{
	Key Laboratory of Modern Astronomy and Astrophysics (Nanjing University), Ministry of Education, Nanjing, Jiangsu 210093, China}
	
\altaffiltext{10}{
	Collaborative Innovation Center of Modern Astronomy and Space Exploration, Nanjing, Jiangsu 210093, China}
	
\altaffiltext{11}{
	International Centre for Radio Astronomy Research, Curtin University, GPO Box U1987, Perth, WA 6845, Australia}
	
\altaffiltext{12}{
	Department of Physics, Drexel University, 3141 Chestnut Street, Philadelphia, PA 19104, USA}
	
\altaffiltext{13}{
	Department of Astronomy and Institute of Theoretical Physics and Astrophysics, Xiamen University, Xiamen, China}

\begin{abstract}
We present \xmm\ imaging spectroscopy of ten weak emission-line quasars (WLQs) at \hbox{$0.928\leq z \leq 3.767$}, six of which are radio quiet and four which are radio intermediate. The new \xray\ data enabled us to measure the power-law photon index, at rest-frame energies $>2$ keV, in each source with relatively high accuracy. These measurements allowed us to confirm previous reports that WLQs have steeper \xray\ spectra, suggesting higher accretion rates with respect to ``typical" quasars. A comparison between the photon indices of our radio-quiet WLQs and those of a control sample of 85 sources shows that the first are significantly higher, at the $\gtsim3 \sigma$ level. Collectively, the four radio-intermediate WLQs have lower photon indices with respect to the six radio-quiet WLQs, as may be expected if the spectra of the first group are contaminated by \xray\ emission from a jet. Therefore, in the absence of significant jet emission along our line of sight, these results are in agreement with the idea that WLQs constitute the extreme high end of the accretion rate distribution in quasars. We detect soft excess emission in our lowest-redshift radio-quiet WLQ, in agreement with previous findings suggesting that the prominence of this feature is associated with a high accretion rate. We have not detected signatures of Compton reflection, \Ka\ lines, or strong variability between two \xray\ epochs in any of our WLQs, which can be attributed to their relatively high luminosity.
\end{abstract}


\section{Introduction}
\label{sec:introduction}
It is common to classify weak emission-line quasars (WLQs) as luminous active galactic nuclei (AGN) having rest frame equivalent widths (EWs) of either $<$15.4\AA\ or $<$10.0\AA\ for the \lya$+$N~{\sc v}~$\lambda1240$ emission complex or C~{\sc iv}~$\lambda1549$ emission line, respectively (Diamond-Stanic \et 2009). These thresholds mark the $3\sigma$ limit at the low-EW tail of the respective EW distributions in Sloan Digital Sky Survey (SDSS; York \et 2000) quasars. Based on this classification, $\approx 10^{3}$ WLQs are known, to date, discovered mainly by the SDSS (e.g., Fan \et 1999; Anderson \et 2001; Collinge \et 2005; Plotkin \et 2010; Meusinger \& Balafkan 2014), but also by other surveys (e.g., McDowell \et 1995; Londish \et 2004). Interestingly, the fraction of WLQs among quasars appears to increase sharply from $\sim0.1\%$ at \hbox{$2 \lesssim z \lesssim 5$} to $\gtrsim15\%$ at $z \gtrsim 6$ (e.g., Fan \et 2006; Diamond-Stanic \et 2009; Ba$\tilde{\rm n}$ados \et 2016). Identifying the cause(s) for their line weakness is therefore important for understanding the physical conditions in the early universe.

Multiwavelength and multi-epoch observations of several sub-samples of WLQs have shown that they are unlikely to be high-redshift galaxies with apparent quasar-like luminosities due to gravitational-lensing amplification, dust-obscured quasars, or broad absorption line (BAL) quasars (e.g., Shemmer \et 2006; Diamond-Stanic \et 2009). Additionally, the radio and \xray\ properties of WLQs indicate that they are unlikely to be identified as high-redshift BL Lacertae objects (Shemmer \et 2009; Plotkin \et 2010; Lane \et 2011). Therefore, the emission lines in WLQs are considered to be intrinsically weak.

Several proposals have been put forward that attempted to explain the intrinsic emission-line weakness in WLQs. One of these suggested that the broad emission line regions (BELRs) in WLQs have either abnormal physical properties (e.g., lack of line-emitting gas, or a low covering factor), or are in the early stages of formation (e.g., Hryniewicz \et 2010; Liu \& Zhang 2011). Although such ideas may appear promising in their attempt to explain the increasing fraction of WLQs as a function of redshift, they face several difficulties, mainly on physical grounds. A different model, suggesting a relatively cold accretion disk, as a result of an unusually high supermassive black-hole mass and low accretion rate (Laor \& Davis 2011), faces its own challenges. In particular, a predicted sharp cutoff in the spectral energy distribution (SED) at $\lambda_{\rm rest}$\ltsim 1000~\AA\ has not yet been detected following observations of several WLQ sub-samples.

The most promising path to identifying the underlying reason for intrinsic BELR line weakness in quasars originates from what is known as the Baldwin effect, which is an anti-correlation between BELR line EW and quasar luminosity (Baldwin 1977). In its modified form, this effect involves an anti-correlation between BELR line EW and the Eddington fraction (i.e., $L$/$L_{\rm Edd}$, where $L$ and $L_{\rm Edd}$ are the bolometric and Eddington luminosity, respectively; e.g., Baskin \& Laor 2004; Dong \et 2009; Shemmer \& Lieber 2015). The idea that a high Eddington fraction, corresponding to a high normalized accretion rate, is responsible for intrinsic line weakness has been proposed in various studies. For example, Leighly \et (2007a,b) have suggested that an extremely high accretion rate would result in a modified, UV-peaked SED lacking high-energy ionizing photons (see also Vasudevan \& Fabian 2007). However, such a model necessarily predicts unusual \xray\ weakness, with respect to the optical emission, which is not observed in all WLQs (e.g., Wu \et 2011, 2012; Luo \et 2015). Quantifying this \xray\ weakness is based on the optical-\xray\ spectral slope, defined as \aox$=\log(f_{\rm 2\,keV}/f_{2500\mbox{\rm\,\scriptsize\AA}})/ \log(\nu_{\rm 2\,keV}/\nu_{2500\mbox{\rm\,\scriptsize\AA}})$, where $f_{\rm 2\,keV}$ and $f_{2500\mbox{\rm~\scriptsize\AA}}$ are the flux densities at 2\,keV and 2500\,\AA, respectively. This parameter is strongly correlated with the luminosity density at 2500~\AA, $L_{\nu}(2500\,{\rm \AA})$ (e.g., Just \et 2007; Lusso \& Risaliti 2016). According to Luo \et (2015), a quasar is considered to be \xray\ weak if it has an observed \aox\ value that is lower by at least 0.2 from the expected value based on the \hbox{\aox-$L_{\nu}(2500\,{\rm \AA})$} correlation, i.e., \hbox{$\Delta$\aox\ $< -0.2$}; otherwise, it is considered \xray\ `normal'.

In order to accommodate the wide range of optical-to-\xray\ flux ratios in WLQs, as well as their other properties, Wu \et (2011) and Luo \et (2015) have proposed an alternative model which also predicts extremely high accretion rates as a primary ingredient to explain quasar emission-line weakness. Unlike the modified SED scenario, this model predicts that the highly ionizing photons are absorbed by a shielding-gas component, growing vertically from the inner accretion disk, perhaps as the accretion rate rises above a certain threshold. This shielding-gas component may be physically identified with the thick inner accretion disk. The range in relative \xray\ weakness is thus explained by a range of viewing angles to the central \xray\ source. When viewed at large inclination angles (i.e., closer to a `pole on' view), a WLQ will appear to have `normal' \xray\ emission with respect to its optical emission; when viewed at smaller inclination angles, a portion of the \xray\ emission is absorbed by the shielding gas, resulting in an \xray\ weak WLQ. This model is supported by observations of \xray\ weak WLQs that show considerably harder \xray\ spectra with respect to typical quasars, indicating heavy intrinsic absorption in such sources (Wu \et 2011, 2012; Luo \et 2015).

However, when compared to typical quasars over wide ranges of redshift and luminosity, WLQs do not appear to follow the strong EW-$L/L_{\rm Edd}$ anti-correlation, where $L/L_{\rm Edd}$ estimates are based on the H$\beta$ line (Shemmer \& Lieber 2015). The EWs of their \civ\ emission lines predict $L/L_{\rm Edd}$ values that are a factor of $\sim5$ larger than estimated. This discrepancy may imply that either (i) there are other factors that regulate emission-line strength in quasars, or (ii) the H$\beta$ line cannot be used to obtain reliable $L/L_{\rm Edd}$ estimates for all quasars. The \xray\ power-law photon index ($\Gamma$), particularly when measured above $\sim2$\,keV in the rest frame, has been identified as a more robust proxy for estimating $L/L_{\rm Edd}$ in quasars (e.g., Shemmer \et 2006, 2008; Constantin \et 2009; Brightman \et 2013; Fanali \et 2013). In particular, Risaliti \et (2009) found a strong correlation (\hbox{$r = 0.56$} and \hbox{$p < 10^{-8}$}) between $\Gamma$ and \hb-based \lledd\ in a sample of 82 SDSS quasars having \xmm\ observations. Accurate measurements of $\Gamma$ in a sizable sample of WLQs can therefore provide an independent indicator of their accretion rates. 

The first steps in this direction were taken by Shemmer \et (2009) and Luo \et (2015) who jointly fitted \xray\ data of seven and 18 WLQs, respectively, obtained from shallow \chandra\ X-ray Observatory (hereafter, \chandra; Weisskopf \et 2000) observations (see also Wu \et 2012). The first of these studies measured a \hbox{$\left < \Gamma \right > = 1.81^{+0.45}_{-0.43}$} in the observed-frame 0.5--8 keV range, concluding that this value is consistent with the values measured in typical type~1 quasars. However, their small sample included a mixture of \xray\ weak and \xray\ normal WLQs with extremely limited photon statistics (hence the large uncertainty on $\left < \Gamma \right >$). The second study measured \hbox{$\left < \Gamma \right > = 2.18\pm{0.09}$} in the rest-frame $>$2 keV band for a well-selected sample of \xray\ normal WLQs with considerably better photon statistics, thereby reducing the uncertainty on $\left < \Gamma \right >$ by a factor of $\approx5$. This recent result shows that \xray\ normal WLQs have, on average, a higher than normal photon index that indicates a high \lledd\ value. It also demonstrates the power of sample averaging in the presence of non-negligible intrinsic scatter that is inherent in the \hbox{$\Gamma$-\lledd} correlation.

In this work we aim to obtain accurate measurements of $\Gamma$ values in a sample of {\em individual} \xray\ normal WLQs in order to determine the extent that these values deviate from the distribution of $\Gamma$ values in typical type~1 quasars. For this purpose, we obtained \xmm\ (Jansen \et 2001) observations of nine high-redshift WLQs discovered by the SDSS that were detected by \chandra. Prior to this investigation, only two such sources were observed by \xmm; one was targeted, and the other was observed serendipitously. We include these two sources in our analysis. We describe our sample selection, observations, and data reduction in Section 2; in Section 3 we present the results from our \xray\ imaging spectroscopy of WLQs and compare them with similar data for a carefully-selected sample of typical quasars. A summary is given in Section 4. Throughout this work we compute luminosity distances using the standard cosmological model (\hbox{$H_{0}$ = 70 km ${\rm s}^{-1}\,\ {\rm Mpc}^{-1}$}, \hbox{$\Omega_{\Lambda}$ = 0.7}, and \hbox{$\Omega_{\rm M}$ = 0.3}; Spergel \et 2007). Complete source names appear in the Tables, and abbreviated names appear in Figures and throughout the text. Unless noted otherwise, hard \xray\ refers to the $>2$ keV energy range in the rest frame, and \nh\ ($N_{\rm H,Gal}$) refers to the intrinsic (Galactic) neutral absorption column density.
\section{Target Selection, Observations, and Data Reduction}
\label{sec:obs}
We selected nine SDSS, \xray\ normal WLQs at \hbox{$0.928\leq z \leq 3.767$} in order of decreasing \xray\ brightness based on previous \chandra\ detections (Shemmer \et 2009; Wu \et 2012; Luo \et 2015). These sources were predicted to provide sufficient \xray\ counts to allow an investigation of their \xray\ spectra with economical \xmm\ observations. As a consequence of our selection algorithm, the two \xray\ brightest targets are also radio intermediate\footnote{The radio loudness parameter, $R$, is defined as $f_{\rm 5\,GHz}/f_{4400\mbox{\rm\,\scriptsize\AA}}$, where $f_{\rm 5\,GHz}$ and $f_{4400\mbox{\rm\,\scriptsize\AA}}$ are the flux densities at 5 GHz and $4400\mbox{\rm\,\scriptsize\AA}$, respectively. Radio quiet (loud) objects are defined as having \hbox{$R<10  ~(R>100)$}.} (\hbox{$10 < R < 100$}; Kellermann \et 1989) assuming a jet is contributing to the radio and \xray\ emissions; the rest are radio quiet. By selection, all of our sources are \xray\ normal (Luo \et 2015) and they have \hbox{$-0.14 < \Delta$\aox $ < +0.35$}.

The \xmm\ observation log appears in \hbox{Table~\ref{tab:obs_log}}. \textit{Column (1)} gives the SDSS quasar name; \textit{Columns (2) and (3)} give the redshift from the SDSS Data Release 7 (Shen \et 2011), and the systemic redshift, respectively; \textit{Column (4)} gives the Galactic absorption column density in units of $10^{20} {\rm cm}^{-2}$, taken from Dickey $\&$ Lockman (1990) and obtained with the HEASARC \nh\ tool\footnote{https://heasarc.gsfc.nasa.gov/cgi-bin/Tools/w3nh/w3nh.pl.}; \textit{Columns (5) and (6)} give the \xmm\ observation ID number and start date, respectively; \textit{Columns (7) - (12)} give the net exposure times and source counts of the MOS1, MOS2, and pn detectors, respectively (these exposure times represent the live time following the removal of flaring periods, and the source counts are in the \hbox{0.2-12.0 keV} band); \textit{Column (13)} gives the radio-loudness parameter (Kellermann \et 1989); \textit{Column (14)} gives the Galactic absorption-corrected flux in the observed-frame $0.5-2$ keV in units of $10^{-15}$ erg cm$^{-2}$ s$^{-1}$ taken from previous \chandra\ data; \textit{Column (15)} gives the reference to the WLQ classification of a source.

Table~\ref{tab:obs_log} includes two additional, radio-intermediate WLQs. The results from an \xmm\ observation of the first of these, SDSS J1141$+$0219 (at \hbox{$z = 3.55$}), have been presented in Shemmer \et (2010); we re-analyze this observation for consistency with the rest of our sample. The second source, SDSS J1012$+$5313 (at \hbox{$z = 2.99$}), has been observed serendipitously by \xmm\ (see below for more details).

\begin{turnpage}
\begin{deluxetable*}{lcccccccccccccc}
\tablecolumns{15}
\tabletypesize{\scriptsize}
\tablewidth{0pc}
\tablecaption{{\sl XMM-Newton} Observation Log \label{tab:obs_log}}
\tablehead
{
\colhead{} &
\colhead{} &
\colhead{} &
\colhead{} &
\colhead{Observation} &
\colhead{Observation} &
\multicolumn{2}{c}{{MOS1}} &
\multicolumn{2}{c}{{MOS2}} &
\multicolumn{2}{c}{{pn}} &
\colhead{} &
\colhead{} \\
\colhead{WLQ} &
\colhead{$z$\tablenotemark{a}} &
\colhead{$z_{\rm sys}$\tablenotemark{b}} &
\colhead{$N_{\rm H,Gal}$\tablenotemark{c}} &
\colhead{{ID}} &
\colhead{Start Date} &
\colhead{Exp Time (ks)} &
\colhead{Counts} &
\colhead{Exp Time (ks)} &
\colhead{Counts} &
\colhead{Exp Time (ks)} &
\colhead{Counts} &
\colhead{$R$\tablenotemark{d}} &
\colhead{$f_{\rm x}$\tablenotemark{e}} &
\colhead{Ref.} \\
\colhead{(1)} &
\colhead{(2)} &
\colhead{(3)} &
\colhead{(4)} &
\colhead{(5)} &
\colhead{(6)} &
\colhead{(7)} &
\colhead{(8)} &
\colhead{(9)} &
\colhead{(10)} &
\colhead{(11)} &
\colhead{(12)} &
\colhead{(13)} &
\colhead{(14)} &
\colhead{(15)}
}
\startdata
\object{SDSS~J090843.25$+$285229.8}\tablenotemark{f} & 0.930 & \nodata & 2.46 & \dataset[]{0760740101} & 2015 Oct 15 & 17.1 & \nodata & 16.9 & \nodata & 12.7 & \nodata & $<2.0$\tablenotemark{i} & 8.30\tablenotemark{i} & 1 \\
\object{SDSS~J092832.87$+$184824.3} & 3.767 & \nodata & 3.83 & \dataset[]{0692510201} & 2012 Oct 24 & 29.3 & 246 & 29.3 & 309 & 24.9 & 996 & 14.7\tablenotemark{j} & 5.17\tablenotemark{j} & 2 \\
\object{SDSS~J101204.04$+$531331.8}\tablenotemark{g} & 2.990 & \nodata & 0.78 & \dataset[]{0651420201} & 2010 Oct 17 & 16.4 & 120 & 16.4 & 84 & 13.1 & 424 & 24.1\tablenotemark{j} & 2.08\tablenotemark{j} & 3 \\
\object{SDSS~J114153.34$+$021924.3} & 3.480 & 3.550\tablenotemark{h} & 2.30 & \dataset[]{0551750301} & 2008 Jun 27 & 20.3 & 162 & 20.3 & 223 & 16.6 & 798 & 14.7\tablenotemark{j} & 2.71\tablenotemark{j} & 4 \\
\object{SDSS~J123132.37$+$013814.0} & 3.229 & \nodata & 1.81 & \dataset[]{0692510101} & 2012 Jun 22 & 19.1 & 270 & 19.1 & 288 & 15.7 & 957 & 39.5\tablenotemark{j} & 7.60\tablenotemark{j} & 4 \\
\object{SDSS~J141141.96$+$140233.9} & 1.745 & 1.754 & 1.43 & \dataset[]{0760740201} & 2015 Jun 30 & 18.5 & 207 & 18.5 & 253 & 15.2 & 700 & $<3.6$\tablenotemark{i} & 5.75\tablenotemark{i} & 1 \\
\object{SDSS~J141730.92$+$073320.7} & 1.704 & 1.716 & 2.12 & \dataset[]{0782360101} & 2017 Jan 4 & 21.3 & 82 & 21.3 & 81 & 17.5 & 265 & $<3.0$\tablenotemark{i} & 3.53\tablenotemark{i} & 1 \\
\object{SDSS~J142943.64$+$385932.2} & 0.928 & \nodata & 0.95 & \dataset[]{0760740401} & 2015 Jun 10 & 25.4 & 166 & 25.4 & 122 & 21.4 & 481 & $<0.9$\tablenotemark{i} & 3.81\tablenotemark{i} & 1 \\
\object{SDSS~J144741.76$-$020339.1} & 1.427 & 1.430 & 4.53 & \dataset[]{0782360201} & 2016 Jul 19 & 36.0 & 223 & 36.0 & 229 & 30.6 & 731 & $<2.3$\tablenotemark{i} & 1.25\tablenotemark{i} & 1 \\
\object{SDSS~J161245.70$+$511816.9} & 1.595 & \nodata & 1.67 & \dataset[]{0743350501} & 2014 Aug 8 & 35.3 & 308 & 35.0 & 385 & 29.8 & 1103 & $<1.5$\tablenotemark{k} & 3.92\tablenotemark{k} & 1 \\
\object{SDSS~J164302.03$+$441422.1}\tablenotemark{l} & 1.650 & \nodata & 1.52 & \dataset[]{0760740301} & 2015 Jun 24 & 17.7 & 357 & 17.4 & 442 & 15.7 & 1183 & $<3.1$\tablenotemark{i} & 5.52\tablenotemark{i} & 1
\enddata
\tablenotetext{a}{Redshift from SDSS Data Release 7 (Shen \et 2011).}
\tablenotetext{b}{Systemic redshift; unless otherwise noted, obtained from Plotkin \et (2015). We use this redshift when available, otherwise we use the redshift from Column (2).}
\tablenotetext{c}{Neutral Galactic absorption column density in units of $10^{20}$\,cm$^{-2}$ obtained from Dickey \& Lockman (1990).}
\tablenotetext{d}{Radio-loudness parameter (Kellermann \et 1989).}
\tablenotetext{e}{Galactic absorption-corrected flux in the observed-frame $0.5-2$ keV in units of $10^{-15}$ erg cm$^{-2}$ s$^{-1}$ taken from previous \chandra\ data.}
\tablenotetext{f}{Observation completely ruined by background flaring.}
\tablenotetext{g}{Serendipitous observation.} 
\tablenotetext{h}{Obtained from Shemmer \et (2010).}
\tablenotetext{i}{Obtained from Luo \et (2015).}
\tablenotetext{j}{Obtained from Shemmer \et (2009).}
\tablenotetext{k}{Obtained from Wu \et (2012).}
\tablenotetext{l}{Observation affected by auto-stack problem; pn experienced Full Scientific Buffer during the whole exposure; see the text for more details.}
\tablerefs{(1) Plotkin \et (2010); (2) Shemmer \et (2009); (3) Schneider \et (2007); (4) Collinge \et (2005).}
\end{deluxetable*}
\end{turnpage}
The data were processed using standard \xmm\ Science Analysis System\footnote{http://xmm.esac.esa.int/sas.} v16.0.0 tasks. All but three objects (discussed below) showed no background flaring activity, and therefore were not filtered in time.

For SDSS J0908$+$2852, the majority of the observation was subject to flaring, and therefore this object has been removed from all our analyses below.

For SDSS J1141$+$0219 and SDSS J1417$+$0733, the event files were filtered in time to remove periods of flaring activity in which the count rates for each MOS (pn) detector exceeded 1.0 (5.0) counts ${\rm s}^{-1}$ for SDSS J1141$+$0219, and 0.5 (1.0) counts ${\rm s}^{-1}$ for SDSS J1417$+$0733. 
The higher thresholds used for SDSS J1141$+$0219 are a consequence of the longer period of flaring activity in this observation. Using a lower threshold, e.g., 0.35 (1.0) counts s$^{-1}$ as used in Shemmer \et (2010), would have resulted in a larger fraction of the observation being discarded.

For all objects except SDSS J1012$+$5313 (discussed below), source counts were extracted from each detector using a circular aperture with $r = 30''$ centered on the source.\footnote{This radius corresponds to $\sim85\%$ of the encircled energy for each detector. See section 3.2.1.1 of the \xmm\ Users Handbook (https://xmm-tools.cosmos.esa.int/external/xmm\_user\_support/documentation/uhb/).} Background counts were extracted from a collection of 3-4 nearby source-free regions that were at least as large as the corresponding source region. 

SDSS J1012$+$5313 is serendipitously detected with an angular offset of 2.845$'$ from the aimpoint. Therefore, we extracted the source counts from a larger circular aperture with $r = 48''$ for the MOS detectors, and $r = 55''$ for the pn detector. These larger regions are expected to capture $\sim90\%$ of the encircled energy. Background counts were extracted as above. This object also has a previous \xmm\ observation, ID 0111100201, which is of low quality and is not useful for our purposes (see Shemmer \et 2009).

We note that the observation of SDSS J1643$+$4414 experienced intervals during which the telemetry allocation for the detector was saturated, either due to a bright source level, or to a high background; however, we do not find any indication that this may have significantly affected the source and associated background event files.

For all objects, the spectrum from each detector was grouped with a minimum of 20 counts per bin, using the High Energy Astrophysics Science Archive Research Center (HEASARC) FTOOLS task {\sc grppha}. The net exposure times (i.e., following the removal of periods of flaring activity) and ungrouped source counts in the $0.2-12.0$ keV observed-frame band are given in Table 1.

We used XSPEC v12.9.1 (Arnaud 1996) to jointly fit the three EPIC detector data sets for each object at rest-frame energies greater than 2 keV with a power-law model and a Galactic-absorption component (i.e., {\sc phabs$*$powerlaw} model in XSPEC), which was kept fixed during the fit, as well as a similar model with an added intrinsic neutral-absorption component (i.e., {\sc phabs$*$zphabs$*$powerlaw} model in XSPEC); we used $\chi^{2}$-statistics for all these fits. For all but one object, the best fits rely on the {\sc phabs$*$powerlaw} absorption model in XSPEC. For SDSS J0928$+$1848, an $F$-test shows that a model including the {\sc zphabs} component provides a better fit (although, as can be seen from Table~\ref{tab:WLQfitresults}, the constraints on the neutral absorption column density in this source are not particularly strong).

\begin{figure*}[t]
\epsscale{0.9}
\plotone{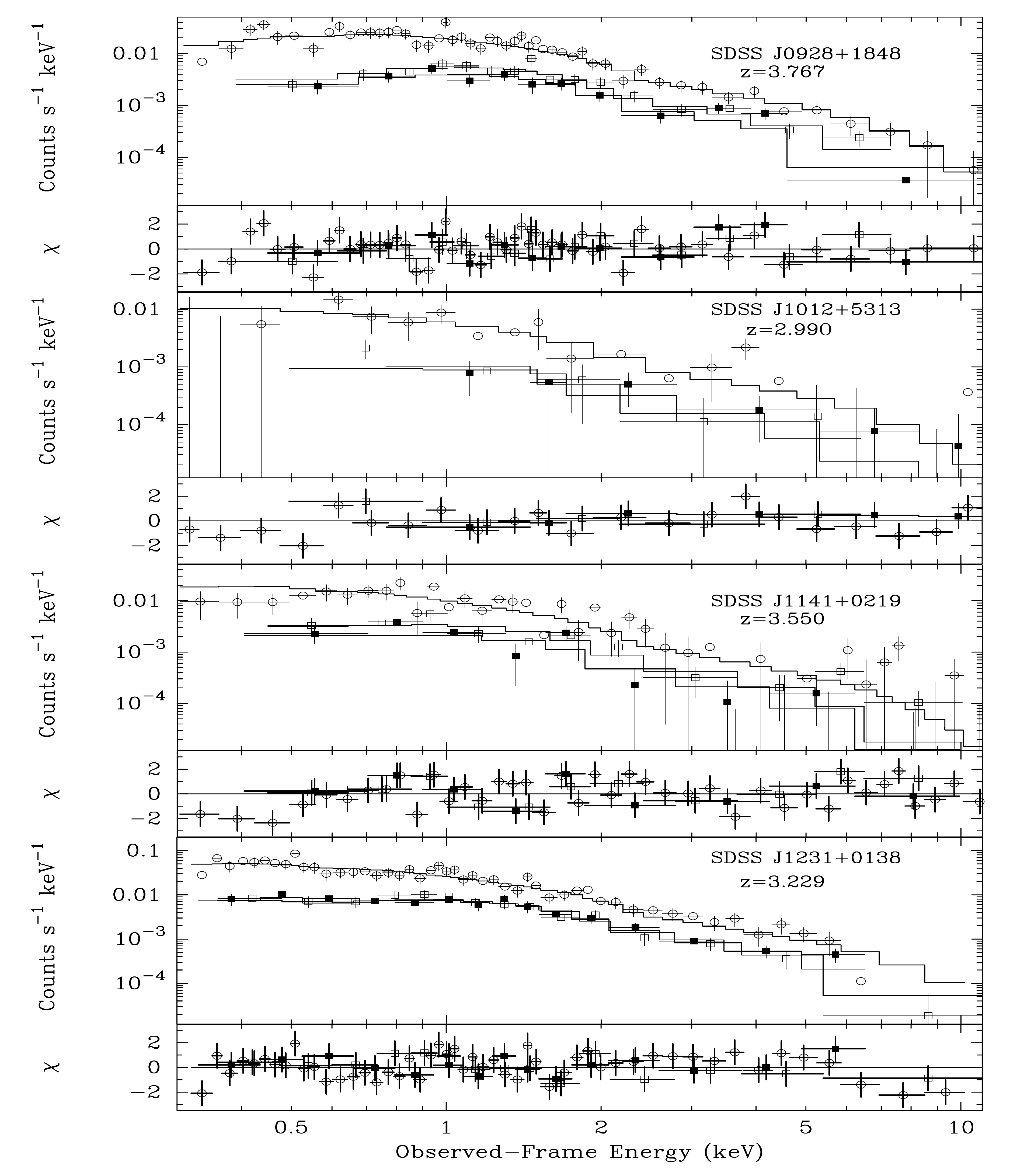}
\caption{Data, best-fit spectra, and residuals of \xmm\ observations of our radio-intermediate WLQs. Open circles, filled squares, and open squares represent the EPIC pn, MOS1, and MOS2, data, respectively; solid lines represent the best-fit model for each spectrum. The data were fitted with a Galactic-absorbed power-law model (with added intrinsic neutral absorption for SDSS J0928$+$1848) above a rest-frame energy of 2 keV, and then extrapolated to 0.3 keV in the observed frame for display purposes. The $\chi$ residuals are in units of $\sigma$ with error bars of size 1.}
\label{fig:radio_int_spec}
\end{figure*}

\begin{figure*}[t]
\epsscale{0.7}
\plotone{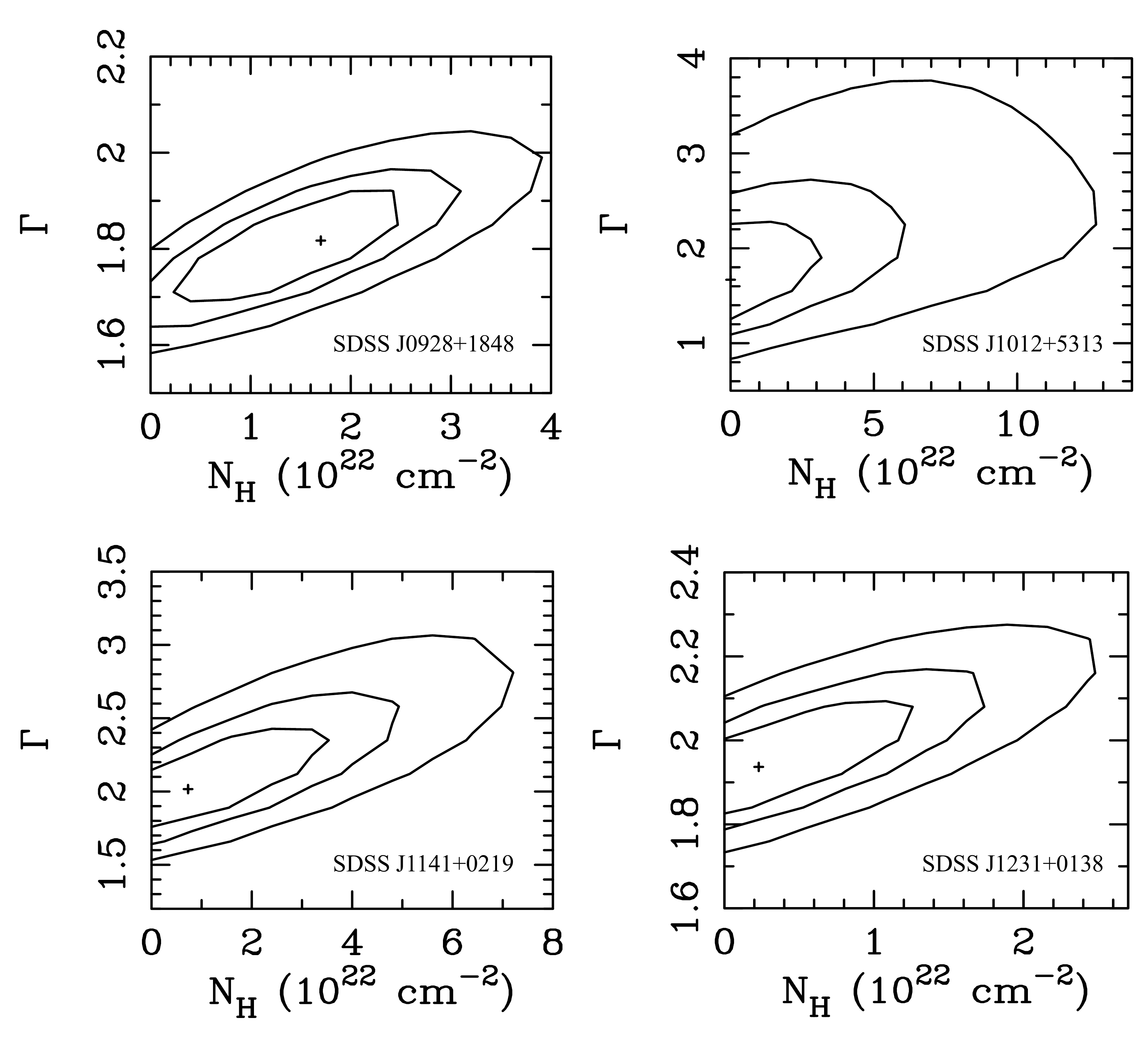}
\caption{The 68\%, 90\%, and 99\% confidence regions for the photon index vs. intrinsic neutral absorption column density of our radio-intermediate WLQs.}
\label{fig:radio_int_contours}
\end{figure*}

\begin{figure*}[t]
\epsscale{0.7}
\plotone{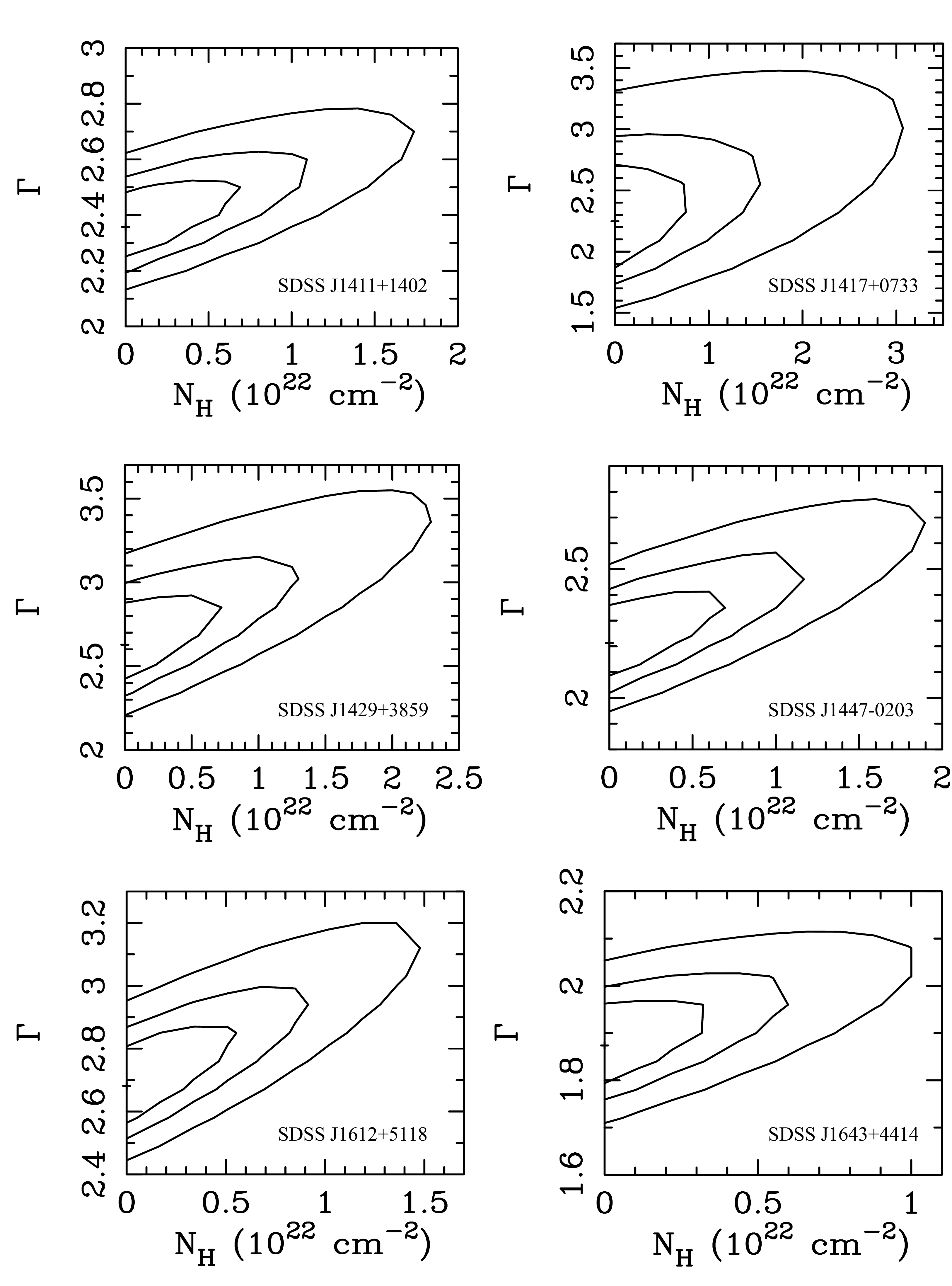}
\caption{Same as Figure~\ref{fig:radio_int_contours} but for our radio-quiet WLQs.}
\label{fig:radio_quiet_contours}
\end{figure*}

The best-fit \xray\ spectral parameters as well as the optical properties of our sample are given in Table~\ref{tab:WLQfitresults}. \textit{Column (1)} gives the SDSS quasar name; \textit{Columns (2) - (4)} give the best-fit $\Gamma$ values, power-law normalizations, and $\chi^{2}$ values, respectively, in the rest-frame $>2$ keV energy range; \textit{Column (5)} shows the upper limits on intrinsic neutral absorption column density (\nh); \textit{Column (6)} gives the monochromatic luminosity at a rest-frame wavelength of 2500\AA\ [$\nu L_{\nu}(2500\,\mbox{\AA})$]; \textit{Columns (7) and (8)} give the \aox\ and the $\Delta$\aox\ parameter, which is the difference between the measured \aox\ and the predicted \aox , based on the \hbox{\aox-$L_{\nu}(2500\,{\rm \AA})$} relation in quasars (given as eq. [3] of Just \et 2007); both parameters are taken from the archival \chandra\ observations of each source. \textit{Column (9)} shows the \aox\ values as measured from our \xmm\ data and the optical data from Column (6); \textit{Column (10)} gives the time separation between the \chandra\ and \xmm\ epochs in the rest-frame; \textit{Column (11)} gives the photon index obtained from fitting the \xray\ spectrum in the observed frame \hbox{$0.5-8$ keV} band (see Section~\ref{sec:L16} for more details). Figures~\ref{fig:radio_int_spec} and \ref{fig:radio_quiet_spec} present the \xmm\ data, their joint, best-fit spectra, and residuals. Correspondingly, Figures~\ref{fig:radio_int_contours} and \ref{fig:radio_quiet_contours} show the 68\%, 90\%, and 99\% confidence regions for the photon index vs. intrinsic neutral absorption column density resulting from those fits when a neutral intrinsic-absorption component is included.

\begin{deluxetable*}{lcccccccccccc}[t]
\tablecolumns{11}
\tabletypesize{\scriptsize}
\tablewidth{0pc}
\tablecaption{Best-Fit X-Ray Spectral Parameters and Optical Properties \label{tab:WLQfitresults}}
\tablehead{
\colhead{} &
\colhead{} &
\colhead{} &
\colhead{} &
\colhead{} &
\colhead{$\log \nu L_{\nu}(2500\,\mbox{\AA})$} &
\multicolumn{2}{c}{{\chandra}} &
\colhead{\xmm} &
\colhead{$\Delta t$\tablenotemark{d}} &
\colhead{} \\
\colhead{{WLQ}} &
\colhead{{\sc $\Gamma$}} &
\colhead{$f_{\nu}$(1\,keV)\tablenotemark{a}} &
\colhead{$\chi^{2}/$(d.o.f.)} &
\colhead{\nh\tablenotemark{b}} &
\colhead{(erg\,s$^{-1}$)} &
\colhead{\aox} &
\colhead{$\Delta$\aox\tablenotemark{c}} &
\colhead{\aox} &
\colhead{(days)} &
\colhead{$\Gamma_{0.5-8~{\rm keV}}$\tablenotemark{e}} \\
\colhead{(1)} &
\colhead{(2)} &
\colhead{(3)} &
\colhead{(4)} &
\colhead{(5)} &
\colhead{(6)} &
\colhead{(7)} &
\colhead{(8)} &
\colhead{(9)} &
\colhead{(10)} &
\colhead{(11)} \\
\noalign{\smallskip}\hline\noalign{\smallskip}
\multicolumn{11}{c}{Radio Intermediate}
}    
\startdata
\object{SDSS~J092832.87$+$184824.3}\tablenotemark{f} & $1.82\pm{0.13}$ & $14.5^{+1.9}_{-1.7}$ & 67/70 & $1.7^{+1.4}_{-1.3}$ & 47.2\tablenotemark{g} & -1.59\tablenotemark{g} & +0.20 & -1.59 & 445 & $1.65^{+0.63}_{-0.26}$ \\
\object{SDSS~J101204.04$+$531331.8} & $1.67^{+0.64}_{-0.48}$ & $3.8\pm{1.1}$ & 18/27 & $\le3.7$ & 46.3\tablenotemark{g} & -1.49\tablenotemark{g} & +0.18 & -1.56 & 869 & $2.14^{+1.02}_{-1.90}$ \\
\object{SDSS~J114153.34$+$021924.3} & $1.93^{+0.26}_{-0.23}$ & $6.8\pm{1.0}$ & 59/54 & $\le3.9$ & 46.7\tablenotemark{g} & -1.54\tablenotemark{g} & +0.18 & -1.55 & 111 & $2.18^{+0.40}_{-1.27}$ \\
\object{SDSS~J123132.37$+$013814.0} & $1.91^{+0.11}_{-0.10}$ & $17.2\pm{1.3}$ & 67/69 & $\le1.4$ & 46.7\tablenotemark{g} & -1.37\tablenotemark{g} & +0.35 & -1.43 & 429 & $1.71^{+0.24}_{-0.33}$ \\
\noalign{\smallskip}\hline\noalign{\smallskip}
\multicolumn{11}{c}{Radio Quiet} \\
\noalign{\smallskip}\hline\noalign{\smallskip}
\object{SDSS~J141141.96$+$140233.9} & $2.36^{+0.14}_{-0.13}$ & $24.0\pm{2.0}$ & 65/51 & $\le0.8$ & 46.0\tablenotemark{h} & -1.42\tablenotemark{h} & +0.20 & -1.35 & 336 & $2.75^{+0.93}_{-0.49}$ \\
\object{SDSS~J141730.92$+$073320.7} & $2.25^{+0.51}_{-0.42}$ & $5.2\pm{1.1}$ & 22/17 & $\le0.9$ & 46.1\tablenotemark{h} & -1.56\tablenotemark{h} & +0.08 & -1.67 & 549 & $2.75^{+0.69}_{-0.38}$ \\
\object{SDSS~J142943.64$+$385932.2} & $2.63^{+0.27}_{-0.25}$ & $19.7^{+3.1}_{-2.8}$ & 28/33 & $\le0.8$ & 45.9\tablenotemark{h} & -1.73\tablenotemark{h} & -0.11 & -1.62 & 474 & $4.51^{+0.39}_{-0.46}$ \\
\object{SDSS~J144741.76$-$020339.1} & $2.21^{+0.16}_{-0.15}$ & $11.9^{+1.3}_{-1.2}$ & 52/53 & $\le0.8$ & 46.0\tablenotemark{h} & -1.76\tablenotemark{h} & -0.14 & -1.55 & 528 & $2.41^{+0.29}_{-0.13}$ \\
\object{SDSS~J161245.70$+$511816.9} & $2.68^{+0.14}_{-0.13}$ & $20.5^{+1.5}_{-1.4}$ & 76/79 & $\le0.6$ & 46.4\tablenotemark{i} & -1.67\tablenotemark{i} & +0.01 & -1.57 & 495 & $2.89^{+0.22}_{-0.18}$ \\
\object{SDSS~J164302.03$+$441422.1} & $1.88\pm{0.10}$ & $32.2\pm{2.2}$ & 86/90 & $\le0.4$ & 46.0\tablenotemark{h} & -1.43\tablenotemark{h} & +0.19 & -1.31 & 285 & $2.04^{+0.20}_{-0.15}$
\enddata
\tablecomments{Unless otherwise noted, the best-fit photon index, normalization, and $\chi^{2}$ were obtained from a Galactic absorbed power-law model. Errors represent 90\% confidence limits, taking one parameter of interest ($\Delta\chi^{2} = 2.71$). The photon index in Column (2) was measured in the rest-frame $>$2 keV energy range.}
\tablenotetext{a}{Power-law normalization given as the flux density at an observed-frame energy of 1\,keV with units of
  10$^{-32}$\,erg\,cm$^{-2}$\,s$^{-1}$\,Hz$^{-1}$; this refers to the pn data and, except for one source, was taken from joint fitting of all three EPIC detectors with the Galactic-absorbed power-law model. The data for SDSS~J092832.87$+$184824.3 was taken with the Galactic-absorbed power-law model with an added neutral intrinsic absorption component.}
\tablenotetext{b}{Intrinsic neutral absorption column density in units of 10$^{22}$\,cm$^{-2}$. Upper limits were computed with the intrinsically absorbed power-law model with Galactic absorption, and represent 90\% confidence limits for each value.}
\tablenotetext{c}{The difference between the measured \aox from column (7) and the predicted \aox based on the \aox-$L_{\nu}(2500\,{\rm \AA})$ relation (given as eq. [3] of Just \et 2007).}
\tablenotetext{d}{Time separation between the \chandra\ and \xmm\ epochs in the rest-frame.}
\tablenotetext{e}{Photon index obtained from fitting the \xray\ spectrum in the observed frame $0.5-8$ keV band using the {\sc pexrav} model in XSPEC (see Section~\ref{sec:L16} for more details).}
\tablenotetext{f}{X-ray spectral parameters were obtained from a Galactic absorbed power-law model that included an intrinsic neutral-absorption component.}
\tablenotetext{g}{Obtained from Shemmer \et (2009).}
\tablenotetext{h}{Obtained from Luo \et (2015).}
\tablenotetext{i}{Obtained from Wu \et (2012).}
\end{deluxetable*}
\begin{figure*}[t]
\epsscale{0.9}
\plotone{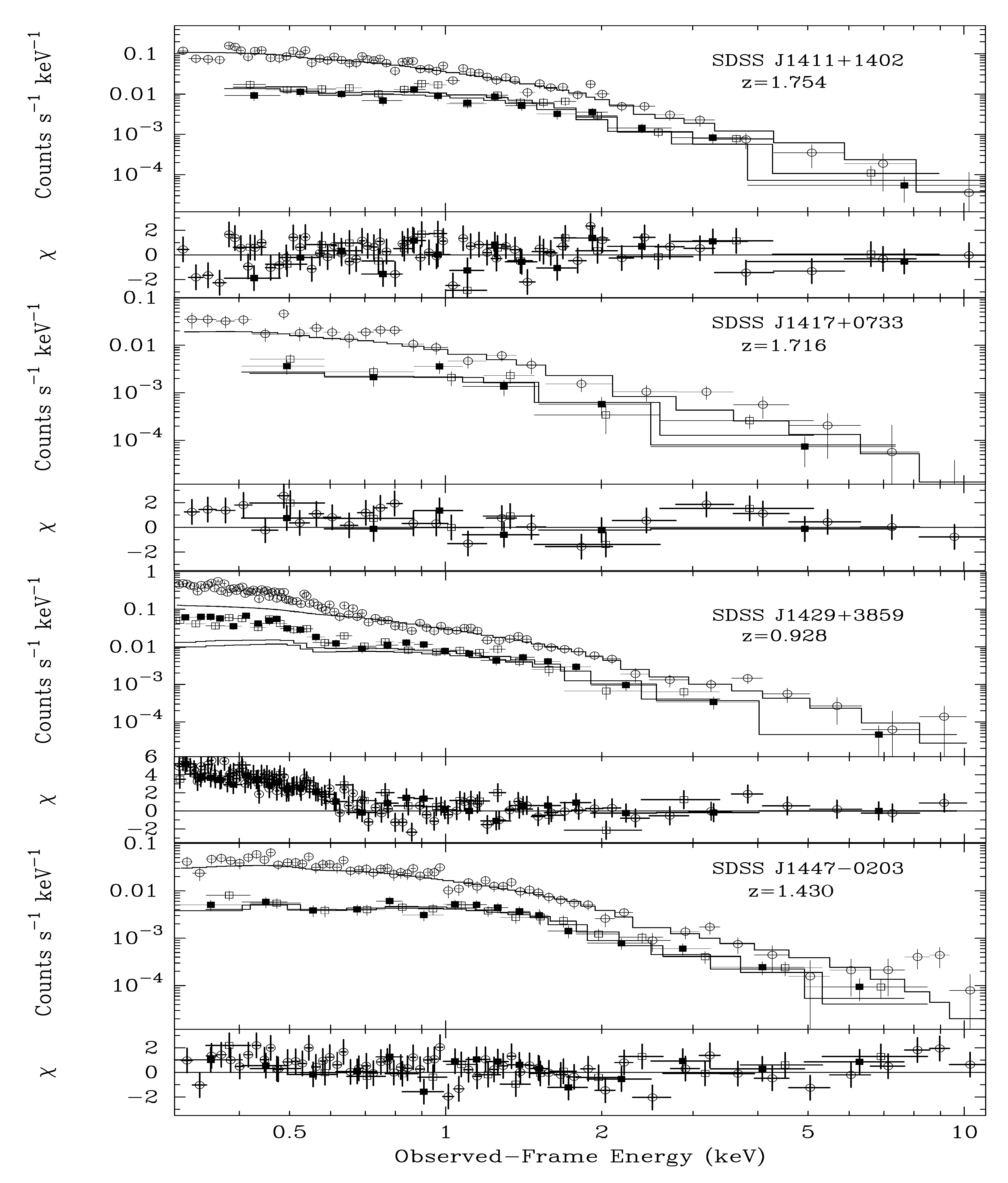}
\caption{Data, best-fit spectra, and residuals of \xmm\ observations of our radio-quiet WLQs. Symbols are similar to those of Figure~\ref{fig:radio_int_spec}.}
\label{fig:radio_quiet_spec}
\end{figure*}
\begin{figure*}[t]
\figurenum{2}
\epsscale{0.9}
\plotone{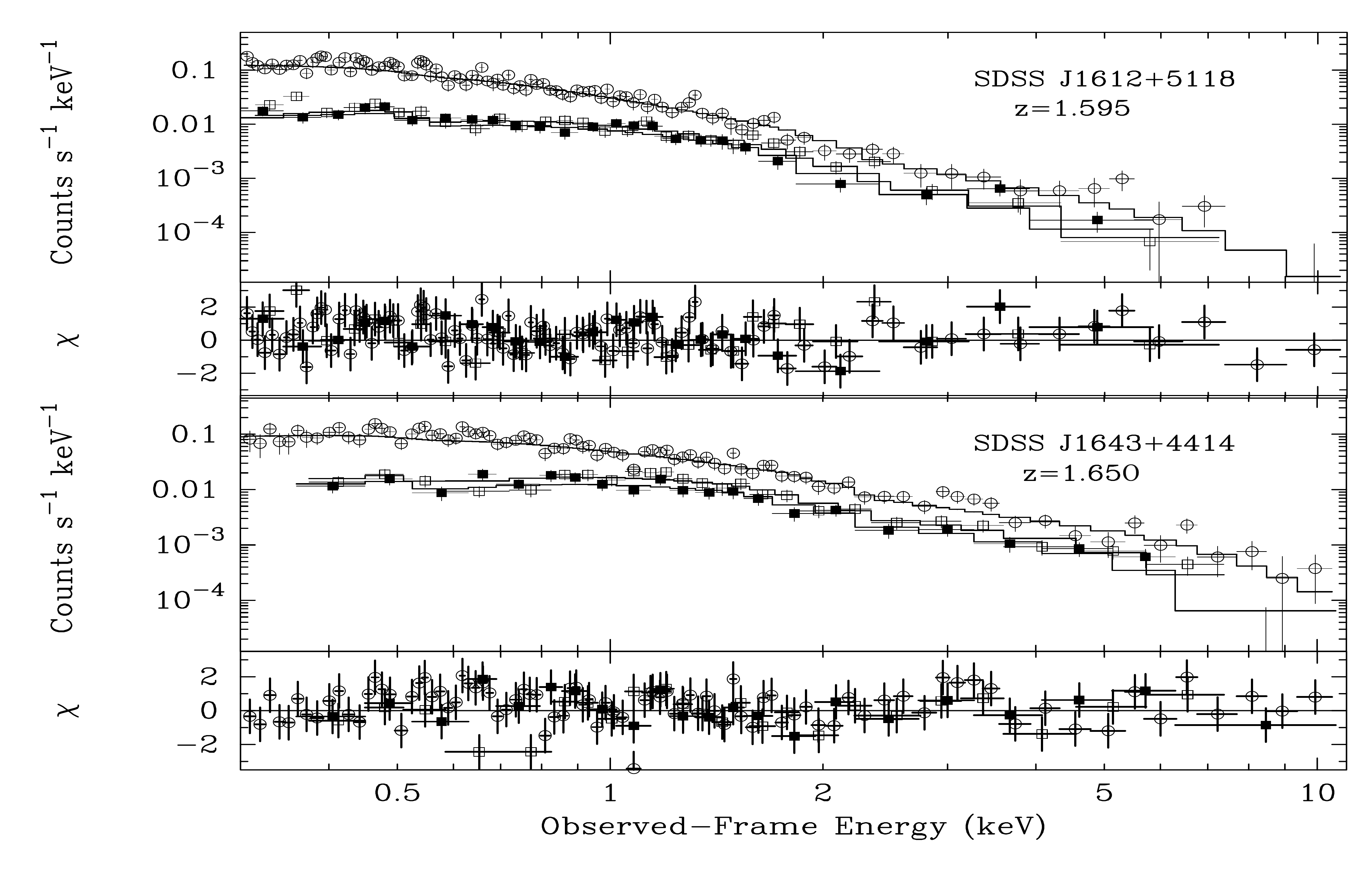}
\caption{Continued.}
\label{fig:radio_quiet_spec_cont}
\end{figure*}

\begin{figure*}[t]
\epsscale{1.0}
\plotone{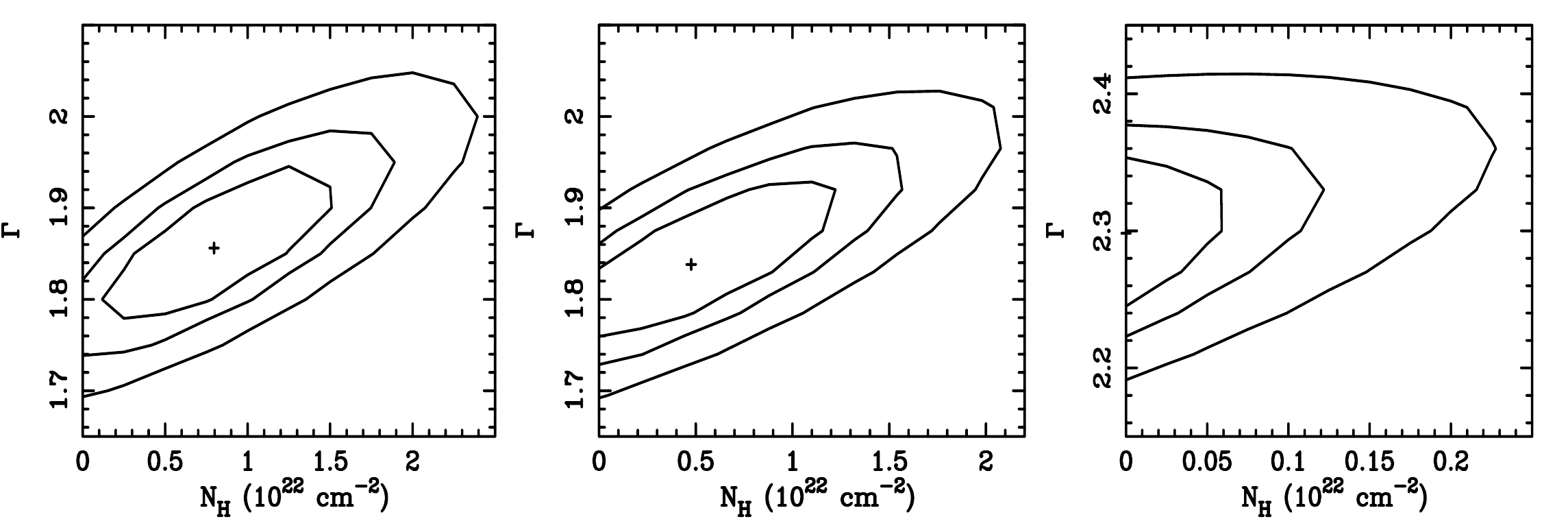}
\caption{The 68\%, 90\%, and 99\% confidence regions for the photon index vs. intrinsic neutral absorption column density derived from the joint spectral fitting of our sample of radio-intermediate (left), radio-intermediate, excluding SDSS~J1012$+$5313 (center), and radio-quiet (right) WLQs.}
\label{fig:contour}
\end{figure*}

In order to obtain better constraints on the $\Gamma$ values of WLQs, as a group, we performed a series of joint spectral fitting of all of our sources in the $>2$ keV rest-frame energy range. Table~\ref{tab:joint_fitting} presents the results of joint-fitting the spectra of all four radio-intermediate sources and all six radio-quiet sources, separately. \textit{Columns (1) and (2)} give the run sequence and number of sources per run, respectively; \textit{Column (3)} gives the mean and median redshifts; \textit{Column (4)} gives the mean intrinsic neutral absorption column density (\nh); \textit{Columns (5) and (6)} give the mean, best-fit $\Gamma$ values and $\chi^{2}$ values, respectively, in the rest-frame $>2$ keV range. The joint fitting was performed three times using the same models as noted above for the individual sources. The first run was completed with objects from Table~\ref{tab:WLQfitresults} that are radio intermediate; the second run is similar to the first, but excluding SDSS J1012$+$5313 which has a relatively lower-quality dataset. The third run included all the objects from Table~\ref{tab:WLQfitresults} that are radio quiet. Figure~\ref{fig:contour} shows the contour plots of the \hbox{$\Gamma$-\nh} parameter space from each joint-fitting run. The results of our joint-fitting show that the radio-quiet and radio-intermediate objects (excluding SDSS~J1012$+$5313) have significantly different mean $\Gamma$ values, with the first group having significantly steeper \xray\ spectra, and there is no detection of significant intrinsic absorption. 

\begin{deluxetable*}{lccccc}[t]
\tablecolumns{6}
\tabletypesize{\scriptsize}
\tablewidth{0pc}
\tablecaption{Best-Fit Parameters from Joint Fitting of \xmm\ Spectra \label{tab:joint_fitting}} 
\tablehead
{
\colhead{} &
\colhead{Number of} &
\colhead{Mean (Median)} &
\colhead{$\left <  N_{\rm H} \right >$} &
\colhead{} &
\colhead{} \\
\colhead{Run}&
\colhead{Sources} &
\colhead{Redshift} &
\colhead{$(10^{22}$~cm$^{-2})$} &
\colhead{\sc $\left < \Gamma \right >$} &
\colhead{$\chi^{2}/$(d.o.f.)} \\
\colhead{(1)} &
\colhead{(2)} &
\colhead{(3)} &
\colhead{(4)} &
\colhead{(5)} &
\colhead{(6)} 
}
\startdata
Radio Intermediate\tablenotemark{a} & 4 & 3.38 (3.39) & $0.80^{+0.84}_{-0.78}$ & $1.86\pm{0.10}$ & 236/231 \\
Radio Intermediate (excluding SDSS~J1012$+$5313) & 3 & 3.52 (3.55) & $\le1.33$ & $1.79\pm{0.06}$ & 214/201 \\
Radio Quiet & 6 & 1.51 (1.62) & $\le0.07$ & $2.30\pm{0.06}$ & 428/349 
\enddata
\tablecomments{Best-fit parameters of joint fitting the spectra in the $>2$ keV rest-frame energy range with a power-law model and Galactic absorption.}
\tablenotetext{a}{X-ray spectral parameters were obtained from a Galactic absorbed power-law model that included an intrinsic neutral-absorption component.}
\end{deluxetable*}

\section{Results and Discussion}
\label{sec:results}
\subsection{How Extreme are the Hard-\xray\ Spectral Slopes of WLQs?}
\label{sec:L16}
Type~1 quasars are known to exhibit a hard-\xray\ spectral slope of $\Gamma \sim 1.8-2.0$ across the Universe (e.g., Reeves \& Turner 2000; Page \et 2005; Piconcelli \et 2005; Shemmer \et 2005; Vignali \et 2005; Just \et 2007; Young \et 2009) that appears to be regulated by \lledd\ (Shemmer \et 2008). 
Based upon the well-known $\Gamma$-\lledd\ correlation, the small fraction of quasars with measured $\Gamma$ values of $\gtsim2.2$ are interpreted as sources that accrete close to or even above the Eddington limit (e.g., Risaliti \et 2009). A natural explanation for the mean $\Gamma$ value of $2.18\pm0.09$, measured for 18 \xray\ normal WLQs by Luo \et (2015) is that these sources lie at the extreme high end of the \lledd\ distribution in quasars. Quantifying, or constraining, their deviations from that distribution can be done by obtaining a deeper \xray\ observation for each individual source.

Our data provide almost an order of magnitude increase in the number of \xray\ counts for radio-quiet and \xray\ normal WLQs, and they confirm the basic Luo \et (2015) finding. Table~\ref{tab:WLQfitresults} shows that most of our radio-quiet sources have extremely high $\Gamma$ values, and the mean $\Gamma$ value of these sources, $\left < \Gamma \right > = 2.30\pm0.06$, based on jointly fitting their spectra (Table~\ref{tab:joint_fitting}) is larger than, yet consistent within the errors with, the Luo \et (2015) value. In order to quantify the extremity of the $\Gamma$ values of these WLQs, we searched the literature and \xray\ archives to identify the most suitable comparison sample of quasars.

\begin{figure*}[t]
\epsscale{0.9}
\plotone{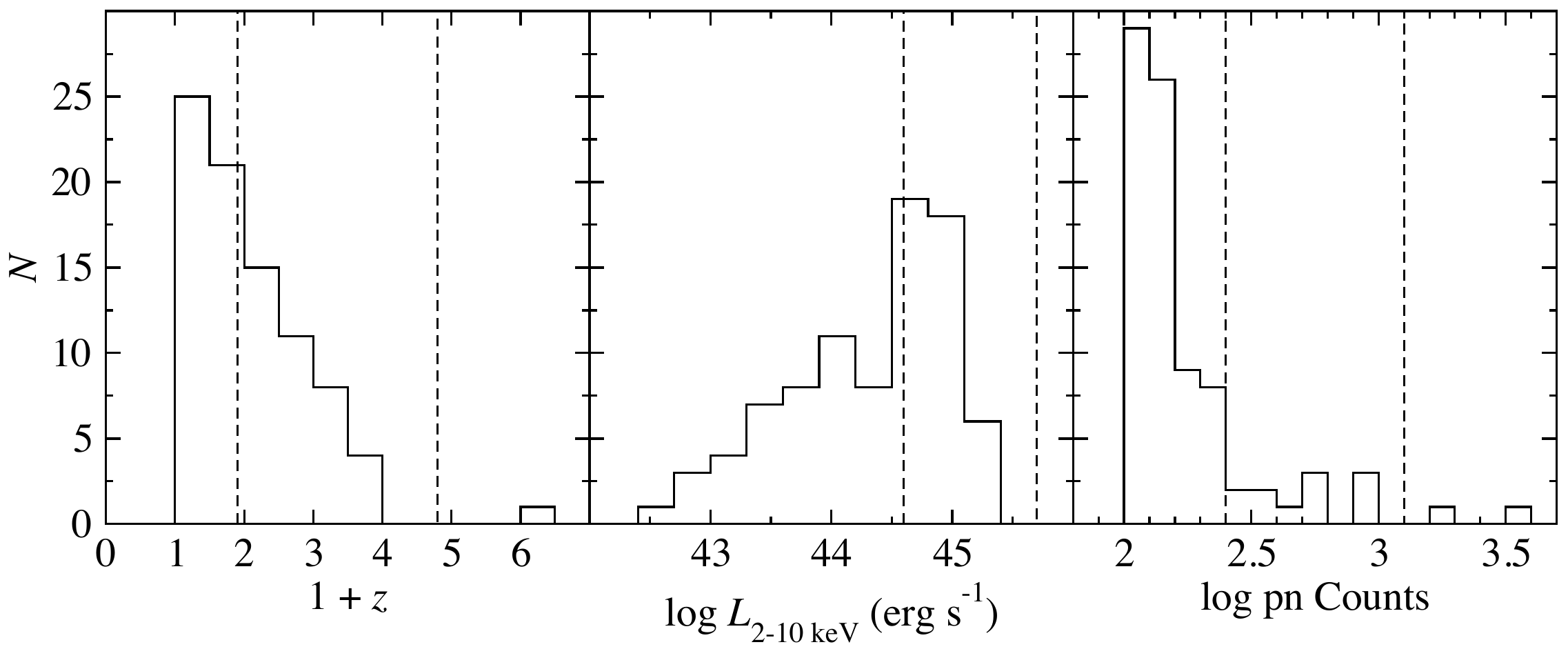}
\caption{Distributions of the redshifts (left), luminosities (middle), and number of pn counts (right) of the L16 comparison sample. Dashed lines represent the upper and lower limits of our WLQ sample in each panel.}
\label{fig:test1}
\end{figure*}

The sample of Liu \et (2016; hereafter, L16) includes 1786 type~1 quasars observed with \xmm\ as part of the XMM-XXL-North survey; this is one of the largest samples of \xray\ detected quasars to date. Of these, 1731 sources are part of the SDSS Data Release 12 quasar catalog (P\^{a}ris \et 2017) which are covered in the Faint Images of the Radio Sky at Twenty cm (FIRST; Becker \et 1995) footprint. We further limited this sample by requiring each source to meet all of the following criteria:
\begin{itemize}
\item{pn counts $> 100$}
 
\item{BAL flag equals zero for sources at $z>1.57$, according to the P\^{a}ris \et (2017) catalog}

\item{radio-quiet sources having $R < 10$}

\end{itemize}
The first of these criteria ensures that sources have \xray\ data with comparable quality to our WLQ observations (a higher pn counts threshold would have limited the sample considerably; see Figure~\ref{fig:test1}). The second criterion is set to minimize the effects of \xray\ absorption (e.g., Gallagher \et 2006), and the third criterion is required for minimizing the potential contribution of a jet to the \xray\ emission (e.g., Miller \et 2011).

There are 167 L16 sources that meet the first criterion, 48 of which are at $z>1.57$. One of these 48 sources is flagged as a BAL quasar by P\^{a}ris \et (2017) and is therefore removed from our sample. Based on the BAL fraction at $z>1.57$, we expect that $\sim 2-3$ BAL quasars may be present among the remaining 119 sources at $z<1.57$. Cross-matching with the FIRST catalog, using a 2$''$ search radius around the SDSS coordinates of each source (see, e.g., P\^{a}ris \et 2017), yielded 15 radio counterparts to the remaining 166 L16 sources. Two additional sources, out of 166, have FIRST detections with angular offsets of 2.2$''$ and 2.7$''$; we consider these to be physically related to the respective L16 sources, thus raising the total number of radio counterparts to 17.

The $R$ values for the radio counterparts were derived by taking the FIRST flux densities at an observed-frame frequency of 1.4 GHz and extrapolating to a rest-frame frequency of 5 GHz, assuming a radio power-law continuum of the form $f_{\nu} \propto \nu^{-0.5}$ (Rector \et 2000). These flux densities were then divided by the flux densities from the $i$-band ($AB_{7672}$) magnitudes taken from the P\^{a}ris \et (2017) catalog and extrapolated to a rest-frame wavelength of $4400\,{\rm \AA}$, assuming an optical power-law continuum of the form $f_{\nu} \propto \nu^{-0.5}$ (Vanden Berk \et 2001). Only two of the 17 radio counterparts were found to be radio quiet; the other 15 were therefore culled from the comparison sample.

For the 149 L16 sources that do not have radio counterparts, we computed upper limits on their $R$ values as described above, except that the radio fluxes were derived by multiplying the RMS radio flux at the SDSS position by a factor of 3. In order to meet our third criterion above and to ensure that only radio-quiet sources are considered for the comparison, we further excluded all sources that have upper limits on $R$ that are greater than 10. The final sample, hereafter the L16 comparison sample, includes 85 sources, 62 of which are at $z<1.57$ (if, instead, we used a more conservative constraint on the radio-undetected sources, by multiplying their RMS fluxes by a factor of 5, this would have reduced the sample size to 63 sources). Assuming the BAL quasar contamination fraction above, we can expect the L16 comparison sample to include not more than $\sim1$ BAL quasar at $z<1.57$. Figure~\ref{fig:test1} presents distributions of the redshift, luminosity, and number of pn counts for the L16 comparison sample.

\begin{figure*}[t]
\epsscale{0.8}
\plotone{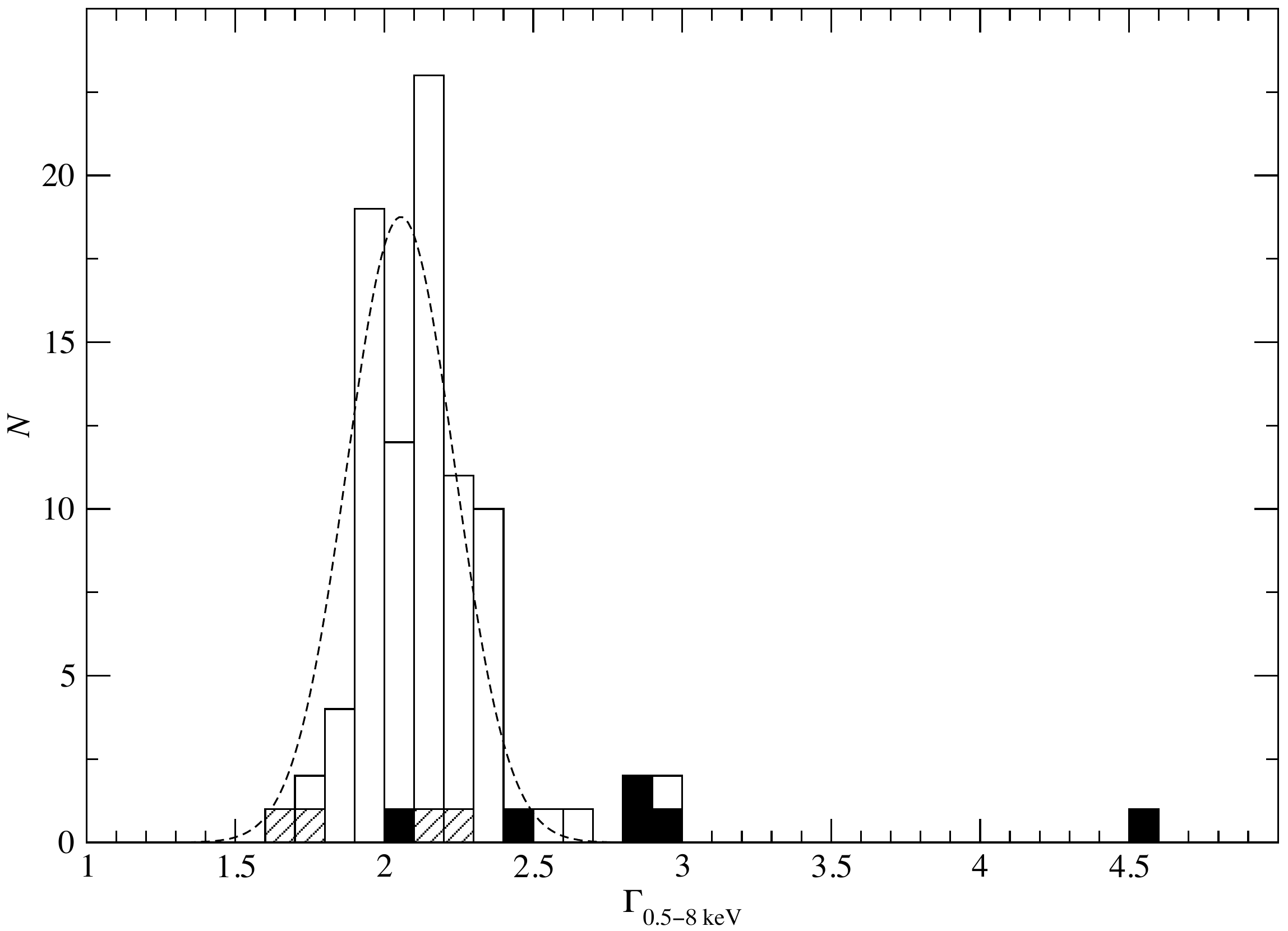}
\caption{Comparison of power-law photon indices measured in the observed-frame $0.5-8$ keV between our WLQs and sources from the L16 comparison sample. The unshaded histogram represents the L16 comparison sample, the dashed curve is the best-fit Gaussian distribution for this histogram, and the hatched and solid black bars represent our radio-intermediate and radio-quiet WLQs, respectively.}
\label{fig:test2}
\end{figure*}

L16 took a Bayesian approach to the \xray\ spectral analysis, using the {\sc bntorus} model.\footnote{This model includes an intrinsic power-law component and Compton scattering from absorbing material. When fitting unobscured quasars like those in our sample, the results of this model are in excellent agreement with those of the {\sc pexrav} model in XSPEC (Magdziarz \& Zdziarski 1995; see Brightman \et 2015 for more details).} Unlike our analysis in the $>2$ keV rest-frame energy range, L16 fit their spectra in the \textit{observed-frame} $0.5-8$ keV range of each source. By adopting this band, the L16 fitting procedure results in non-uniform sampling of their \xray\ spectra, given the wide range of source redshifts. One effect that may stem from employing this procedure is the measurement of an unrealistically high $\Gamma$ value. As we discuss below, this is most likely a consequence of including a soft-excess component in the spectral fitting. The distribution of the $\Gamma_{0.5-8~{\rm keV}}$ values, as measured by L16, for the L16 comparison sample appears in Figure~\ref{fig:test2}.
In order to obtain a meaningful comparison between the L16 $\Gamma_{0.5-8~{\rm keV}}$ values and the $\Gamma$ values of our WLQs, we re-fitted each of our objects in the observed-frame $0.5-8$ keV range, this time with the Galactic-absorbed power-law and Compton reflection ({\sc phabs$*$pexrav}) model (a similar model with an additional intrinsic absorption component, {\sc zphabs}, was used for SDSS J0928$+$1848; see Section~\ref{sec:obs}). We ran {\sc pexrav} while fixing only the redshifts and the Galactic absorptions; all the other model parameters were free to vary. The best-fit $\Gamma_{0.5-8~{\rm keV}}$ values resulting from these fits\footnote{The corresponding best-fit relative Compton-reflection parameters have not yielded meaningful constraints on the Compton-reflection component in any of our WLQs, similar to results we present in Section~\ref{sec:compt_reflect} below.} appear in Column (11) of Table~\ref{tab:WLQfitresults}. The $\Gamma_{0.5-8~{\rm keV}}=4.51$ value for SDSS J1429$+$3859 appears to be unphysical, but can be explained by the indication of excess soft-\xray\ emission at \hbox{$<$ 2 keV} in the rest-frame (see Figure~\ref{fig:radio_quiet_spec}).

Furthermore, we searched for a potential systematic offset between the method L16 used to measure their $\Gamma_{0.5-8~{\rm keV}}$ values and the analysis method we used to obtain the $\Gamma_{0.5-8~{\rm keV}}$ values of our WLQs. Therefore, we have reanalyzed the \xmm\ spectra of seven sources from the L16 comparison sample that had greater than 600 total counts per source. Such a threshold on the number of counts ensures that we compare L16 data sets of roughly matched quality to those of our radio-quiet WLQs (see Table~\ref{tab:obs_log}).
We employed the same data reduction and analysis as described above on single L16 data sets of those seven sources; since each L16 source typically has 1--10 exposures per pointing with a range of angular offsets from the aimpoint, we used the data set with the longest exposure time in each case. As done in L16, we restricted the fitting range to 0.5--8 keV in the observed frame of each source, then fitted each data set once with a Galactic-absorbed power-law and Compton reflection ({\sc phabs$*$pexrav}) model and a second time with an added intrinsic absorption component {\sc (phabs$*$zphabs$*$pexrav}); based on $F$-tests, none of the spectra warranted a neutral intrinsic absorption component. The results of this analysis are given in Table~\ref{tab:L16_seven}. \textit{Column (1)} gives the \xray\ source identification string used by L16; \textit{Columns (2) and (3)} give the SDSS quasar name and corresponding \xmm\ observation ID number, respectively; \textit{Columns (4) and (5)} give the redshifts and number of pn counts taken from L16, respectively; \textit{Columns (6) and (7)} give the $\Gamma_{0.5-8~{\rm keV}}$ values from L16 and our {\sc pexrav} analysis, respectively.
We found no significant systematic difference between the L16 $\Gamma_{0.5-8~{\rm keV}}$ values and the $\Gamma_{0.5-8~{\rm keV}}$ values we obtained for these seven sources.\footnote{We also analyzed the data set of the source from the L16 comparison sample with the largest value of $\Gamma_{0.5-8~{\rm keV}}$ (SDSS J022039.48$-$030820.3 with \hbox{$\Gamma_{0.5-8~{\rm keV}}= 2.91^{+0.08}_{-0.20}$}) in the same way and found a $\Gamma_{0.5-8~{\rm keV}}$ value of $2.66^{+0.18}_{-0.17}$. However, we note that this data set contains only 106 pn counts, and the source is detected close to the edge of the detector.}

Similar to the analysis described in Section~\ref{sec:obs} for our WLQs, we also fitted those seven L16 sources with a Galactic-absorbed power-law model and found a systematic offset of $\approx +0.2$ between $\Gamma_{0.5-8~{\rm keV}}$({\sc pexrav}) and $\Gamma_{0.5-8~{\rm keV}}$(power-law), as may be expected, given that the {\sc pexrav} model attempts to include an additional component due to reflection from neutral material.
Additionally, we note that the seven L16 sources with the largest number of counts are not identical to those with the highest $\Gamma_{0.5-8~{\rm keV}}$ values. Therefore, our results are not biased by sources with the steepest spectral slopes. 
Figure~\ref{fig:test2} shows how the $\Gamma_{0.5-8~{\rm keV}}$ values of our WLQs compare with those of the L16 comparison sample. 

\begin{deluxetable*}{lcccccc}[t]
\tablecolumns{7}
\tabletypesize{\scriptsize}
\tablewidth{0pc}
\tablecaption{Best-Fit Parameters of L16 Sources With Largest Number of Counts \label{tab:L16_seven}} 
\tablehead
{
\colhead{} &
\colhead{} &
\colhead{Observation} &
\colhead{} &
\colhead{} &
\colhead{{\sc $\Gamma_{0.5-8~{\rm keV}}$}} &
\colhead{{\sc $\Gamma_{0.5-8~{\rm keV}}$}} \\
\colhead{UXID}&
\colhead{SDSS Name} &
\colhead{ID} &
\colhead{$z$} &
\colhead{pn Counts} &
\colhead{(from L16)} &
\colhead{(from {\sc pexrav})} \\
\colhead{(1)} &
\colhead{(2)} &
\colhead{(3)} &
\colhead{(4)} &
\colhead{(5)} &
\colhead{(6)} &
\colhead{(7)} 
}
\startdata
N\_38\_68 & \object{SDSS~J021808.24$-$045845.2} & 0112371001 & 0.714 & 3308 & $2.27\pm{0.02}$ & $2.53\pm{0.09}$ \\
N\_38\_117 & \object{SDSS~J021817.45$-$045112.5} & 0112371001 & 1.083 & 1985 & $1.95\pm{0.03}$ & $1.88^{+0.09}_{-0.17}$ \\
N\_20\_50 & \object{SDSS~J022105.64$-$044101.5} & 0037982001 & 0.199 & 974 & $2.11\pm{0.03}$ & $2.16^{+0.09}_{-0.15}$ \\
N\_0\_30 & \object{SDSS~J022224.20$-$034757.3} & 0604280101 & 1.687 & 938 & $1.87^{+0.07}_{-0.06}$ & $1.97^{+0.19}_{-0.27}$ \\
N\_27\_19 & \object{SDSS~J022244.40$-$043347.0} & 0109520601 & 0.761 & 849 & $2.15\pm{0.05}$ & $2.30^{+0.16}_{-0.14}$ \\
N\_113\_13 & \object{SDSS~J022851.50$-$051223.1} & 0677590132 & 0.316 & 619 & $2.13\pm{0.04}$ & $2.17^{+0.12}_{-0.08}$ \\
N\_38\_79 & \object{SDSS~J021830.59$-$045622.9} & 0112371001 & 1.397 & 611 & $2.28\pm{0.05}$ & $2.63^{+0.49}_{-0.35}$ 
\enddata
\end{deluxetable*}

The $\Gamma_{0.5-8~{\rm keV}}$ values of four of our WLQs, all of which are radio-quiet, lie above the $3\sigma$ threshold at the high end of the $\Gamma_{0.5-8~{\rm keV}}$ distribution of the L16 comparison sample; similarly, the $\Gamma_{0.5-8~{\rm keV}}$ value of another radio-quiet WLQ lies at the $\sim2 \sigma$ threshold. Importantly, the average $\Gamma_{0.5-8~{\rm keV}}$ value of our six radio-quiet WLQs ($\Gamma_{0.5-8~{\rm keV}} = 2.89^{+0.45}_{-0.30}$) also lies above the $3\sigma$ threshold of the $\Gamma_{0.5-8~{\rm keV}}$ distribution of the L16 comparison sample (this average drops to
\hbox{$\Gamma_{0.5-8~{\rm keV}} = 2.28^{+0.51}_{-0.57}$} when SDSS J1429$+$3859 is excluded). 

In order to check whether the $\Gamma_{0.5-8~{\rm keV}}$ values of our WLQs, as a group, are significantly higher than those of typical quasars, we ran a Mann-Whitney nonparametric rank test between the $\Gamma_{0.5-8~{\rm keV}}$ values of our six radio-quiet WLQs and those of the L16 comparison sample. We found that the two distributions are significantly different, with $>99.8\%$ confidence ($> 3\sigma$), one-tailed (another test with the exclusion of SDSS J1429$+$3859 resulted in the two distributions being significantly different with $>99.5\%$ confidence). We also ran a similar Mann-Whitney test between the $\Gamma_{0.5-8~{\rm keV}}$ distributions of our six radio-quiet and four radio-intermediate WLQs, and similarly found that the two distributions are significantly different at the $95\%$ confidence level. This result is also reflected in Table~\ref{tab:joint_fitting}. The lower $\Gamma$ values of the radio-intermediate WLQs, with respect to their radio-quiet counterparts, may be a manifestation of jet contributions to their \xray\ emissions (e.g., Miller \et 2011).

Our results therefore indicate that, in the absence of potential \xray\ emission from a jet along our line of sight, WLQs have significantly higher hard \xray\ power-law photon indices than typical quasars. This result reinforces the idea that weak emission lines in quasars may be a direct consequence of a high Eddington fraction. In this respect, our results are in agreement with the Luo \et (2015) model, which suggests that the scale height of the inner accretion disk grows as a function of the accretion rate and acts as a filter that prevents highly ionizing photons from reaching the BELR. However, in order to establish a relationship between BELR line strength and the Eddington fraction across wide ranges of these parameters, the hard \xray\ power-law photon indices of a statistically meaningful sample of quasars should be measured accurately (see, e.g., Shemmer \& Lieber 2015).

\subsection{A Soft Excess - Accretion Rate Connection?}
\label{sec:soft_excess}
The discrepancies in the $\Gamma$ values of our WLQs between those fitted in the $>2$ keV rest-frame band and those in the observed-frame $0.5-8$ keV band (see Table~\ref{tab:WLQfitresults}) could be attributed to the existence of soft \xray\ excess emission, at least in our lowest-redshift sources. The physical nature of this component is uncertain (Porquet \et 2004; Gierlinski \& Done 2004; Vasudevan \et 2014), yet it is present in many AGN spectra which makes it of interest to search for the existence of this component in our sources. In order to check whether any of our sources shows evidence for a soft excess, we extrapolated the best-fit Galactic absorbed power-law model (with added intrinsic neutral absorption for SDSS J0928$+$1848) obtained for rest-frame energies $>2$ keV (see Section~\ref{sec:obs} and Table~\ref{tab:WLQfitresults}) to the \hbox{$>0.3$ keV} observed-frame energies. All but one of our sources show $\chi$ residuals no greater than the $3\sigma$ level and, therefore, no indication of excess soft-\xray\ emission. Only one of our WLQs, SDSS J1429$+$3859, has an indication of soft excess emission with $\chi$ residuals up to $6\sigma$ (see Figure~\ref{fig:radio_quiet_spec}). This is our lowest-redshift WLQ. The non-detection of this feature in the other WLQs is not unexpected given their considerably higher redshifts and the $\sim0.2$ keV energy threshold of \xmm\ (see Shemmer \et 2008).

In order to assess the effect of the putative soft excess on the photon index of SDSS J1429$+$3859, we performed an additional spectral fitting on this source in which a thermal component (the {\sc nlapec} model in {\sc XSPEC}) was added to the model employed in Section~\ref{sec:L16} (i.e., {\sc phabs$*$pexrav$+$nlapec}).
This fitting resulted in a photon index value of \hbox{$\Gamma_{0.5-8~{\rm keV}} = 2.57^{+0.39}_{-0.68}$}.
An $F$-test shows that the addition of the {\sc nlapec} component provides a significantly better fit, with $>90$\% confidence, and that the $\Gamma_{0.5-8~{\rm keV}}$ value is reduced considerably with respect to the one from Section~\ref{sec:L16} ($\Delta \Gamma_{0.5-8~{\rm keV}} \simeq 2$).

A soft excess feature is expected to be more pronounced in sources with higher accretion rates (e.g., Done \et 2012). The fact that we detect a feature of this kind in one of our radio-quiet sources is in agreement with the idea that WLQs have extremely high accretion rates. However, \xray\ imaging spectroscopy of additional WLQs is required to establish such a connection.

\subsection{Searching for Signatures of Compton Reflection and Iron-Line Emission}
\label{sec:compt_reflect}
We conducted a search for the existence of a Compton-reflection continuum as well as signatures of a neutral narrow \Ka\ emission line at rest-frame 6.4 keV in our WLQs in order to assess their potential effects on our photon index measurements. This was performed by fitting all the \xmm\ spectra for each source in the $>2$ keV rest-frame energy range with XSPEC, employing a Galactic absorbed power-law with a Compton-reflection continuum model (i.e., the {\sc pexrav} model in XSPEC, using a similar spectral fitting approach as the one performed in Section~\ref{sec:L16}), and a Galactic absorbed power-law with a redshifted Gaussian emission line model ({\sc zgauss} in XSPEC) for the \Ka\ emission line. The Gaussian rest-frame energy and width were fixed at \hbox{$E = 6.4$ keV} and \hbox{$\sigma = 0.1$ keV}, respectively. Table~\ref{tab:EW} lists the best-fit parameters from these fits. \textit{Column (1)} gives the SDSS quasar name; \textit{Column (2)} gives the rest-frame EW of the \Ka\ emission line; \textit{Column (3)} gives the relative-reflection component ($R_{\rm rel}$) of the Compton-reflection continuum expressed as \hbox{$R_{\rm rel} = \Omega/2\pi$}, where $\Omega$ is the solid angle subtended by the continuum source. Due to the relatively low quality of the SDSS J1012$+$5313 observation, this object was not included in this portion of the analysis. 

\begin{deluxetable*}{lcc}[t]
\tablecolumns{3}
\tabletypesize{\scriptsize}
\tablewidth{0pt}
\tablecaption{Compton Reflection and Iron Emission \label{tab:EW}}
\tablehead
{
\colhead{} &
\colhead{EW(\Ka)\tablenotemark{a}} &
\colhead{} \\
\colhead{WLQ} &
\colhead{(eV)} &
\colhead{$R_{\rm rel}$\tablenotemark{b}} \\
\colhead{(1)} &
\colhead{(2)} &
\colhead{(3)} \\
\noalign{\smallskip}\hline\noalign{\smallskip}
\multicolumn{3}{c}{Radio Intermediate}
}
\startdata
\object{SDSS~J092832.87$+$184824.3}\tablenotemark{c} & $\le319$ & $\le5.8$ \\
\object{SDSS~J101204.04$+$531331.8} & \nodata & \nodata \\
\object{SDSS~J114153.34$+$021924.3} & $\le531$ & $\le4.6$ \\
\object{SDSS~J123132.37$+$013814.0} & $\le62$ & $\le6.1$  \\
\noalign{\smallskip}\hline\noalign{\smallskip}
\multicolumn{3}{c}{Radio Quiet} \\
\noalign{\smallskip}\hline\noalign{\smallskip}
\object{SDSS~J141141.96$+$140233.9} & $\le121$ & $\le96.7$ \\
\object{SDSS~J141730.92$+$073320.7} & $\le326$ & $\le50.1$ \\
\object{SDSS~J142943.64$+$385932.2} & $\le90$ & $\le163.5$ \\
\object{SDSS~J144741.76$-$020339.1} & $\le42$ & $\le12.3$ \\
\object{SDSS~J161245.70$+$511816.9} & $\le170$ & $\le9.6$ \\
\object{SDSS~J164302.03$+$441422.1} & $\le160$ & $\le3.4$ 
\enddata
\tablecomments{Best-fit parameters of fitting each spectrum at the $>2$ keV rest-frame energy range with a model consisting of a Galactic absorbed power-law, a Compton-reflection component, and a neutral \Ka\ emission line.}
\tablenotetext{a}{Rest-frame equivalent width of a neutral \Ka\ emission line at rest-frame $E = 6.4$ keV and a fixed width of $\sigma = 0.1$ keV.}
\tablenotetext{b}{Relative Compton-reflection parameter.}
\tablenotetext{c}{For this object we also included, in addition to the model above, an intrinsic neutral-absorption component.}
\end{deluxetable*}

Previous studies have shown trends where the EW of the narrow \Ka\ line decreases with \xray\ luminosity, i.e. the ``\xray\ Baldwin effect" (e.g., Iwasawa \& Taniguchi 1993; Ricci \et 2013), the origin of which is still unclear. Table~\ref{tab:EW} shows that we have not detected any statistically significant Compton-reflection continua nor any neutral \Ka\ emission in any of our sources. Due to the relatively high luminosities of our sources, these results are in agreement with the \xray\ Baldwin effect. However, our results can be referred to as not being sensitive enough to either confirm or rule out an \xray\ Baldwin Effect. Detection of (or placement of meaningful constraints on) narrow iron lines in such luminous quasars require considerably longer exposures with \xmm. 

\subsection{X-ray Variability}
\label{sec:variability}
Table~\ref{tab:WLQfitresults} shows that no unusual \xray\ variations are observed for any of our WLQs between their \chandra\ and \xmm\ epochs, separated by $\approx 1$ yr in the rest frame of each source. The differences between each pair of \aox\ values indicates \xray\ variations of up to a factor of $\sim3.5$ (in SDSS J1447$-$0203), which is consistent with the \xray\ variability of typical quasars having similar luminosities (e.g., Vagnetti \et 2013; Lanzuisi \et 2014). Therefore, we do not expect that our main results are affected by \xray\ variability. 
\section{Summary}
\label{sec:summary}
We present \xray\ spectroscopy of ten SDSS, \xray\ normal WLQs at \hbox{$0.928\leq z \leq 3.767$} that have sufficient \xray\ counts to allow basic measurements of their \xray\ spectra with \xmm\ observations. Six of these are radio-quiet and four are radio-intermediate. Our analysis provides measurements of the hard \xray\ photon index in these sources. We have compared these data with similar data for a carefully-selected sample of 85 radio-quiet type 1 quasars in order to quantify the extremity of the hard-\xray\ spectral slopes of WLQs with respect to typical quasars. The results of this comparison show that the photon indices of radio-quiet WLQs, as a group, constitute the $>3\sigma$ tail of the photon index distribution in quasars. The radio-intermediate WLQs have considerably lower photon indices which are comparable to those of the bulk of the quasar population; we suggest that \xray\ emission from a jet contributes to the harder \xray\ spectra in these sources. Considering the hard \xray\ power-law photon index as an Eddington-fraction indicator, our results imply that radio-quiet WLQs occupy the extreme high end of the accretion-rate distribution in quasars.

Our lowest-redshift radio-quiet WLQ, SDSS J1429$+$3859, is our only source that exhibits soft excess emission which may be another manifestation of its high accretion rate. None of our sources shows signatures of Compton reflection, or the presence of a narrow iron line, and none shows unusual \xray\ variability. These results are in line with the high luminosity of our sources.

In the near future, a more rigorous comparison with the $\Gamma$ values of our WLQs could be made using recent and deeper \xmm\ observations of a subset of the L16 comparison sample (Chen \et 2018). The new spectra of these sources will be fitted also at the $>2$ keV rest-frame range to conform with the spectral fitting of our WLQs. Sensitive \xray\ imaging spectroscopy of a large sample of quasars across a wide range of BELR line strength, redshift, and luminosity, is required for establishing connections between BELR line strength, Eddington fraction, and the prominence of a soft excess component in all quasars.
\acknowledgements
We gratefully acknowledge the financial support of NASA grants NNX13AB57G, NNX16AC06G, and NNX17AC67G (A.~M., O.~S.),
the NASA ADAP program (W.~N.~B.), Curtin University through the Peter Curran Memorial Fellowship (R.~M.~P.), the National Key R \& D Program of China grant 2016YFA0400702 and National Natural Science Foundation of China grant 11673010 (B.~L.).
Support for this work was also provided by the National Aeronautics and Space Administration through Chandra Award Number AR3-14009X and GO7-18110X issued by the Chandra X-ray Center, which is operated by the Smithsonian Astrophysical Observatory for and on behalf of the National Aeronautics and Space Administration under contract NAS8-03060 (G.~T.~R).
We thank an anonymous referee for a thoughtful and constructive report that helped in improving this manuscript.
This research has made use of the NASA/IPAC Extragalactic Database (NED) which is operated by the Jet Propulsion Laboratory, California Institute of Technology, under contract with the National Aeronautics and Space Administration, as well as NASA's Astrophysics Data System Bibliographic Services.
This research has also made use of data provided by the High Energy Astrophysics Science Archive Research Center (HEASARC), which is a service of the Astrophysics Science Division at NASA/GSFC and the High Energy Astrophysics Division of the Smithsonian Astrophysical Observatory.

\end{document}